\documentclass[12pt]{article}
\usepackage{latexsym}
\usepackage{epsfig,amssymb,euscript}
\usepackage{amsmath}
\usepackage{array,calc,epsfig}

\def\be{\begin{equation}}
\def\ee{\end{equation}}
\def\bea{\begin{eqnarray}}
\def\eea{\end{eqnarray}}
\usepackage[]{graphicx}

\def\tr           {\mathop{\rm Tr}}

\def\sqr#1#2{{\vcenter{\vbox{\hrule height.#2pt  
 \hbox{\vrule width.#2pt height#1pt \kern#1pt \vrule width.#2pt}\hrule  
 height.#2pt}}}}

\begin{document}

\begin{titlepage}

\font\cmss=cmss10 \font\cmsss=cmss10 at 7pt
\leftline{\tt hep-th/0503242}

\vskip -0.5cm
\rightline{\small{\tt KUL-TF-04/40}}
%\rightline{\small{\tt CPHT-RR-114-ssss}}
%\rightline{\small{\tt IFUM-yyy-FT}}
%\rightline{\small{\tt IAS nnn }}

\vskip .7 cm
%\hfill IC/2004/ \vskip .1in \hfill CPHT \vskip .1in \hfill hep-th/0503242

\hfill
\vspace{18pt}
\begin{center}
{\Large \textbf{Semiclassical strings on confining backgrounds}}
\end{center}

\vspace{6pt}
\begin{center}
{\large\textsl{Luca Martucci}} 

\vspace{25pt}
\textit{Institute for Theoretical Physics, K.U. Leuven,\\ Celestijnenlaan 200D, B-3001 Leuven, Belgium}\\  \vspace{4pt}
and\\  \vspace{4pt}
\textit{Dipartimento di Fisica, Universit\`a di Milano, \\ Via Celoria, 16; I-20133 Milano, Italy} 
\end{center}

%\textit{$^a$ The Abdus Salam ICTP, Strada Costiera, 11; I-34014 Trieste, Italy.}\\
%\textit{$^b$ Centre de Physique Th\'eorique, \`Ecole Polytechnique, 48 Route de Saclay; F-91128 Palaiseau Cedex, France.}\\
%\textit{$^c$ INFN, Piazza dei Caprettari, 70; I-00186  Roma, Italy.}\\
%\textit{$^d$ Institute for theoretical physics, K.U. Leuven,
%Celestijnenlaan 200D, B-3001 Leuven, Belgium.}
%\end{center}
\vspace{12pt}

\begin{center}
\textbf{Abstract}
\end{center}

\vspace{4pt} {\small \noindent
This paper discusses some results on semiclassical string configurations lying in the IR sector of supergravity backgrounds dual to  confining gauge theories. On  the gauge  theory side, the string states we analyse correspond to Wilson loops, glueballs and KK-hadrons. The 1-loop correction to the classical energy is never subleading,  but can be viewed as a coupling dependent rescaling of the dimensionful parameters of the theory.} 
\vspace{5cm}

\noindent {\em Talk given at the RTN Workshop on the Quantum Structure of Spacetime and the Geometric Nature of Fundamental Interactions, Kolymbari, Crete, Greece, 5-10 Sep 2004.}

\vfill
\vskip 5.mm
\hrule width 5.cm
\vskip 2.mm
{\small
\noindent e-mail: luca.martucci@fys.kuleuven.ac.be}

\end{titlepage}

%%%%%%%%%%%%%%%%%%%%%%%%%%%%%%%%%%%%%%%%
\section{Introduction}
In the last few years semiclassical strings have proved to be a powerful tool to investigate the gauge/string duality beyond the supergarvity approximation. Berenstein, Maldacena and Nastase (BMN) \cite{bmn}  studied the solvable string theory on the plane-wave obtained by a Penrose limit of the $AdS_5\times S^5$ background, extending in this way the check of the standard AdS/CFT duality beyond the supergravity approximation. The BMN approach was revisited by Gubser, Klebanov and Polyakov (GKP) \cite{gkp}, who substituted the large spin/charge limit, automatic in the Penrose limit, by a large spin limit performed directly on solitonic classical strings configurations. In this way, a larger variety of string solitons were considered, differing substantially from the collapsed string corresponding to the light-cone string vacuum on the plane-wave of \cite{bmn}. The GKP approach was extended to multispinning strings in a series of subsequent papers (see the  review article \cite{tseyrew} and references therein), providing several examples of ``regular'' classical string configurations with an energy which admits, as for the string excitations on the plane wave of \cite{bmn}, a regular expansion in the 't Hooft coupling $\lambda=g_sN$. Many of these classical results have been checked by computations of the anomalous dimension of the corresponding operators of the dual ${\cal N}=4$ SYM theory and  a strict relation with the theory of integrable systems has emerged on both sides of the duality (see e.g. \cite{tseyrew,beisert} and references therein).

An extension of obvious interest of the above approach is to study  less supersymmetric and nonconformal cases (see \cite{confrev} for reviews). This paper is devoted to report on work  in collaboration with F. Bigazzi, A. L. Cotrone and L. A. Pando Zayas  \cite{tomilu1,tomilu2} (for other papers on semiclassical strings on confining backgrounds, see e.g. \cite{crucco}). In these papers we have considered semiclassical string states living in the infrared sector of some supergravity backgrounds dual to confining gauge theories, namely the Maldacena-N\`u\~nez (MN) solution \cite{mn} and its softly broken deformation (bMN) \cite{bMN}, and  Witten's background dual to a nonsupersymmetric YM theory \cite{wym}. The Klebanov-Strassler (KS) solution \cite{ks} and its softly broken deformation (bKS) \cite{borgubs,sonne5} is analogous to the (b)MN case.  The common feature of the confining backgrounds  is that the effective geometry relevant for studying classical string configuration in the deep IR sector (i.e. at minimal transverse radius) factorizes in the simplified form $R^{1,3}\times S^q\times R^{6-q}$, where $q=3$ for the (b)MN background  and $q=4$ in the Witten's background. As a consequence, the search for classical string states lying at the minimal radius simplifies considerably. Such factorization is strictly valid only at the minimal radius, due to warping and twisting effects, and then the study of the quantum corrections to these configurations is more complicated. Nevertheless in some particular limit (like the large spin limit), the leading contribution can sometimes be computed exactly. We have done this for open straight strings  and folded closed spinning strings extending in the flat directions, and closed strings extending and spinning on the internal sphere. The first configuration is a good approximation of the relevant contribution to the dual Wilson loop when the two probe quarks are largely separated; the second configurations give rise to glueball Regge trajectories; the third configurations are dual to hadronic states, whose constituents are  massive KK matter which are present in these models and cannot be decoupled in the supergravity approximation.    

Each of the following sections will be devoted (skipping all the technical details) to the discussion of the results  obtained in \cite{tomilu1,tomilu2} for each of the above configurations in turn. In particular, we will see how  in all the cases considered the leading quantum corrections to the classical energy are nonvanishing, mainly due to the absence of any kind of effective supersymmetry on the world-sheet. Nevertheless, the corrections are finite  without the need for any renormalization prescription once the contribution of the fermions is carefully taken into account. In particular,  quantum corrections to the Wilson loop provide (beside the L\"uscher term already considered in \cite{olesen,sonne}) a rescaling of the field theory string tension, consistently with some general arguments presented in \cite{gross}. Glueballs quantum corrections give a rescaling of the QCD string tension consistent with the Wilson loop analysis and render the  Regge trajectory nonlinear with a positive intercept, thus sharing common features with the experimental results (analogous to those discussed in \cite{regge} for the MN and KS models). Finally, the quantum corrections to the energy of closed strings spinning along internal directions have been studied  explicitly for a particular class of  closed circular strings analogous to those considered in \cite{frolov3} in the $AdS_5\times S^5$ case. In our  confining cases, the leading quantum corrections are nonvanishing but can be absorbed in a redefinition of the mass of the basic KK fields which are the constituents of the dual hadrons.

%%%%%%%%%%%%%%%%%%%%%%%%%%%%%%%%%%%%%%%%%%%%%%%%%%%%%%%%%%%
%%%%%%%%%%%%%%%%%%%%%%%%%%%%%%%%%%%%%%%%%%%%%%%%%%%%%%%%%%%

\section{Wilson loops}
The extension of the description of the Wilson loop in standard AdS/CFT duality \cite{malda} to  confing backgrounds has been studied in several papers at the classical and 1-loop level (see for example respectively \cite{wym,clsw} and \cite{olesen,sonne}). In \cite{tomilu2} we considered the problem of computing 1-loop quantum corrections to the classical result in  Witten's model. This background is obtained by considering the near-horizon limit of the supergravity solution corresponding to $N$ D4-branes wrapped on a circle of radius $R_\theta$, with anti-periodic boundary conditions for the fermions. This is  given by 
\bea  
\label{defns} 
ds^2&=&(\frac{u}{R})^{3/2} (\eta_{\mu\nu}dx^\mu dx^\nu + \frac{4R^3}{9u_0}f(u)d\theta^2)+ (\frac{R}{u})^{3/2}\frac{du^2}{f(u)}
+R^{3/2}u^{1/2}d\Omega_4^2\ ,\nonumber \\   e^\Phi&=&g_s\frac{u^{3/4}}{R^{3/4}}\ ,\qquad \qquad \qquad F_{(4)}=3R^3\omega_4\ ,\nonumber\\
f(u)&=&1-\frac{u_0^3}{u^3}\ , \qquad \qquad \qquad R=(\pi Ng_s)^\frac{1}{3}{\alpha'}^\frac{1}{2}\ ,
\eea
where $\omega_4$ is the volume form of the
transverse $S^4$. 
The geometry is completely regular and consists of a warped, flat 4-d part, a radial direction $u$, a circle parameterized by $\theta$ with vanishing radius at the horizon $u=u_0$, and a four-sphere whose volume is instead everywhere non-zero. The complete Wilson loop configuration is given by a solution of the form 
\be
\label{wilgen} t=\tau\ ,\qquad x^1=\sigma\ , \qquad u=u(\sigma)\ , \sigma\in[-\frac L2,\frac L2]\ .  
\ee
As discussed in \cite{olesen,sonne}, this is well approximated by a bathtub shape  in the large $L$ limit, with the string coming down from infinity practically straight up to a minimal $u_m$, suddenly (but smoothly) becoming flat along a special direction ($x^1$ in (\ref{wilgen})), and then returning back to infinity with another straight line.  This string spans the loop traveling for a time $\bar{t}$.  The important point is that the deviation of this smooth configuration from the rectangular well is exponentially vanishing with $L$.  The lines coming down from infinity give the infinite quark masses which require renormalization.  Moreover, in the regime $L \rightarrow \infty$,  one has that $u_m \rightarrow u_0$. This makes the following straight line approximation of the above solution 
\be\label{lineu}   
t=\tau\ ,\qquad x^1=\sigma\
,\quad \sigma\in[-\frac{L}{2},\frac{L}{2}]\ , \qquad u=u_0\ , 
\ee an accurate one\footnote{The approximation of a straight string lying at a constant nonminimal radius $u_m>u_0$ is not reliable, mainly because it does not satisfy the equation of motion. As discussed in \cite{sonne}, such a configuration would give one more ``Goldstone'' massless mode respect to the approximation (\ref{lineu}). But, as shown in appendix C of \cite{tomilu2}, in the large $L$ limit the explicit nontrivial $\sigma$ dependence, that the string configuration must have in order to satisfy the equations of motion when $u_m> u_0$, gives rise to a mass for the would be ``Goldstone''  boson. This mass is equal to the mass obtained from approximation (\ref{lineu}) up to subleading terms.}.      
The classical value of the energy satisfies the area law typical of confinement 
\be 
E = T_{QCD}L\ , \qquad
T_{QCD}=\frac{u_0^{3/2}}{2\pi \alpha'R^{3/2}}=\frac{1}{6\pi}\lambda
m_0^2\ ,
\label{areal}
\ee 
where $m_0\sim 1/R_\theta$ is the KK mass scale and $\lambda=g_{YM}^2N$ is the bare 't Hooft coupling \cite{tomilu2,gross}. Since the supergravity approximation is valid when $\lambda\gg 1$  we can see how in this approximation the fundamental scale of the gauge theory defined by the string tension is much higher than the KK mass scale. Then, as usual in these cases, the KK matter cannot be thought as decoupled from the pure YM content. 

In order to study the 1-loop quantum correction to this formula one needs the explicit  superstring action up to quadratic order in the fluctuations around this configuration. The bosonic action is the usual Polyakov action, while the fermionic quadratic action can be found for example in \cite{luca}. After $\kappa$-fixing the resulting physical fluctuations are given by six massless plus two massive bosonic fields of mass $m_B=3m_0/2$, and by eight massive fermionic fields of mass $m_F=3m_0/4$. The mass matching condition 
\bea\label{massmatch}
\sum_B m_B^2=\sum_F m_F^2\ 
\eea 
then insures that the theory is UV finite and without Weyl anomaly at one loop (see \cite{grosstsey} and appendix D of \cite{tomilu2}).   
The  resulting 1-loop correction to the energy
\be   
E_1=\frac{\pi}{2L}\sum_{n\geq 1}\left[6\sqrt{n^2}+2\sqrt{n^2+\frac{9m_0^2L^2}{4\pi^2}}-8\sqrt{n^2+\frac{9m_0^2L^2}{16\pi^2}}\right]\ ,   
\ee   
is finite without any renormalization prescription and  in the large $L$ limit reads
\be   
E_1\simeq -\frac{9m_0^2L}{8\pi}\log{2} -\frac{\pi}{4L} .
\ee   
The leading term in this expression comes from approximating the series above by integrals. The L\"uscher-like term $\pi/4L$ comes from the subleading contribution of the six massless modes. The remaining subleading terms are exponentially suppressed in the large $L$ limit and are related to the massive modes (see \cite{aldofra,alfra2} for details on the evaluation of series like the one above). Thus in the large $L$ limit the 1-loop corrected energy of
the string configuration corresponding to the Wilson loop is given by 
\be\label{eloopfin}
E\simeq T^{(ren)}_{QCD}(\lambda)L-\frac{\pi}{4L}  \ ,  \ee
with a rescaled $\lambda$ dependent QCD string tension $T^{(ren)}_{QCD}(\lambda)=(1- \frac{27}{4\lambda}\log2)T_{QCD}$.   Then the quantum corrections do not only give the usual L\"uscher term but even changes the $\lambda$ dependence in the  QCD string tension, which is completely consistent with the general argument of \cite{gross}. In fact, one expects that the QCD string tension  is a function  of  $\lambda$ of the form $T^{(ren)}_{QCD}(\lambda)=f(\lambda)m_0^2$, interpolating between the strong coupling behavior $f(\lambda)\sim \lambda$ given in (\ref{areal}) for $\lambda\gg 1$ and the expected behavior $f(\lambda)\sim \exp{(-b/\lambda)}$ (where $b$ is a constant given by the first coefficient of the beta-function) for $\lambda\ll 1$ (the regime in which the KK matter decouples).  See section 6.1 of \cite{tomilu2} for further discussions on this point.

\section{Glueball Regge trajectories}
In confining backgrounds high spin glueballs are naturally associated to closed strings spinning in the flat directions. Due to the warping of the flat spacetime subsector these string configurations naturally live at the minimal transverse radius. Then, a spinning folded string configuration is given by
\bea\label{reggesolu} 
m_0 x^0=k\tau\ ,\qquad
m_0 x^1=k\cos\tau\sin\sigma\ ,\qquad m_0 x^2=k\sin\tau\sin\sigma\ , 
\eea 
where the KK mass scale $m_0$ is considered as the natural scale for all these theories. The energy is given in general by the expected Regge-like relation
\be \label{reggeclass}
E^2 = 2\pi T_{glue}J\quad ,\quad   T_{glue}= 2 T_{QCD}\ .
\ee 

In \cite{tomilu2} the 1-loop quantum corrections to (\ref{reggeclass}) have been discussed  in the case of the Witten model. Since the technical details are quite complicated, we just skip them and quote the final results. The 1-loop  correction to the energy is again finite without the need for any renormalization prescription. This is due to a $\sigma$ dependent  mass-matching condition analogous to (\ref{massmatch}), which ensures the absence of Weyl anomaly on the worldsheet. In particular it has been possible to explicitly compute the 1-loop corrections in the two opposite limits $\lambda\gg J$ and $\lambda\ll J$. The analysis leads to the conclusion that 1-loop quantum corrections  give a deviation from the linear Regge trajectory with a nonzero intercept and  rescale (in an energy dependent way) the effective string tension $ T_{glue}$ appearing in (\ref{reggesolu}). More explicitly, for $J\ll \lambda$, we have obtained
\bea
\label{rt}
J=\alpha^\prime_{glue}(\lambda)\left[E^2 +2E\,z_0+ z_0^2\right] \quad,\quad \alpha^\prime_{glue}(\lambda)^{-1}=T^{(ren)}_{glue}(\lambda)=2T_{QCD}\Big(1-\frac{6z_1}{\lambda\,m_0}\Big)\ ,
\eea
with $z_0\simeq 0.85 m_0$ and $z_1\simeq 0.012$. 
In the opposite high spin regime $J\gg \lambda$, the leading contribution of the one-loop correction rescales the glueball string tension to $T^{(ren)}_{glue}(\lambda)=2T_{QCD}^{(ren)}(\lambda)$. This is completely consistent with the Wilson loop result since at high spin the glueball is usually thought  as a folded and spinning  closed flux-tube configuration. The quantum corrections to this kind of configuration were discussed previously in the context of the (type IIB) MN and KS backgrounds in \cite{regge}. In this paper the explicit calculation was performed in the limit  $J\ll \lambda$ and the approach is similar  although  technically  slightly different from that used in \cite{tomilu2} \footnote{For example in \cite{tomilu2}, being in type IIA, it has not been possible to adopt a covariant $\kappa$-fixing condition analogous to that used in \cite{regge}.}.  As a result, the authors of \cite{regge} found a deviation  from the linear Regge trajectory analogous to that given in (\ref{rt}), but no rescaling of the QCD string tension (see  \cite{regge} also for further discussions on the comparison of these stringy results with lattice and  experimental data).

 Regge trajectories for mesons in the Witten model have been studied at the classical level in the recent paper \cite{regge2}. Flavored fundamental quarks can be introduced by studying probe D6-branes in this background \cite{mateos}; the high spin mesons are identified with spinning  open strings ending on these branes. Let us notice that, as shown in \cite{regge2},  in the large spin limit  the relevant contribution to the energy of these spinning strings come from an almost flat horizontal piece lying very close to the minimal radius of the geometry, the mass of the quarks being negligible. In this limit the 1-loop correction to the classical energy found in \cite{regge2} is then given by one-half the 1-loop correction computed in \cite{tomilu2} for the glueballs. The resulting quantum corrected slope of the Regge trajectory is then given by the inverse of $T_{QCD}^{(ren)}(\lambda)$, again consistent with effective picture of the meson as a spinning open string consisting of two quarks connected by a flux tube.

\section{KK Hadrons}
In the   Witten model and the other confining gauge/gravity  models such as the (b)MN  (and the (b)KS), there are always extra KK matter fields which, in the supergravity approximation we are working in, cannot decouple from the pure (S)YM theory field content. In particular, there are complex  scalar fields $\Phi$ in the adjoint representation that can form hadronic bound states of the symbolic form $\tr \Phi^J|0\rangle_{gauge}$. These states carry $J$ units of the fundamental charge associated to $\Phi$, which is in turn  associated to some internal symmetry of the background. In the Witten  model we have an $SO(5)$ symmetry group of the gauge theory, which corresponds to the isometry group of the internal $S^4$. On the other hand,  in the (b)KS we have the isometry group $SO(4)$ of the internal $S^3$, while in the (b)MN solution this isometry group is broken to $U(1)_1\times U(1)_2$ by twisting effects. In all these cases we can consider closed strings spinning along the internal Killing directions with given angular momenta $J_1,J_2$ with respect to the maximal abelian (sub)group  $U(1)_1\times U(1)_2$. The associated gauge theory states should have the symbolic form
\bea
\tr [\Phi_1^{J_1} \Phi_2^{J_2}]|0\rangle_{\rm gauge}\ ,
\eea
where $\Phi_1$ and $\Phi_2$ are the simplest KK massive scalar fields with the correct mass and quantum numbers in order to match with the corresponding string states.

This has been done for the first time in \cite{pandostras}, where the authors  performed a Penrose limit around a null geodesic moving  along one of the above isometries for the KS and MN solution. In this way they obtained a soluble string theory and predicted   the mass spectrum of hadronic states of the above kind with $J_2\ll J_1$ and very large $J_1$ (for generalization of this BMN-like approach to other cases  see \cite{abc,sonne5,alfra2,tomilu2}). In \cite{tomilu1,tomilu2} we have shown  how to generalize this BMN approach to extended multispinning strings, extending the predicted mass spectrum to states with $J_2$ of the some order of $J_1$. The key point is that, in searching for classical string configurations lying at the minimal radius, we can always focus on an effective geometry of the form $R^{1,3}\times S^3$ for any of these cases. Then, we can borrow all the solitonic solutions already discussed in the case of $AdS_5\times S^5$  which correspond to strings at rest in $AdS_5$ and spinning  on a $S^3\subset S^5$ (see e.g. the review article \cite{tseyrew}). We have considered the explicit case of multi-spinning folded and circular strings which are particular examples of solutions called ``regular'' in  \cite{tseyrew}. In $AdS_5\times S^5$ these solutions admit a regular expansion in the 't Hooft coupling of the dual ${\cal N}=4$ SYM theory and it is believed that their quantum corrections to the classical energy are subleading in the infinite $J$ limit. In our case these corresponds to  strings admitting a classical energy/spins relation of the following form
\bea\label{expa1}
E=m_0J\Big[1+a_1\frac{\lambda^2}{J_1^2}+ b_1 \frac{\lambda^2}{J_2^2} + a_2\Big(\frac{\lambda^2}{J_1^2}\Big)^2 + 
b_2\Big(\frac{\lambda^2}{J_2^2}\Big)^2+\ldots\Big]\ .
\eea 
In \cite{tomilu1}, by an explicit calculation, we have shown  that, differently from the $AdS_5\times S^5$, in the (b)MN background   one-loop corrections  are not in general subleading. Of course, this result can be  adapted to the other confining backgrounds. For example, in the  Witten model  we are explicitly considering in this paper, the leading one-loop correction to the energy of  multispinning circular strings, analogous to those considered in \cite{frolov3}, is given by
\bea
\Delta E_{\rm 1-loop}=m_0J\Big(\frac{45}{8\lambda}\log\frac34\Big)\ ,
\eea 
where $J=J_1+J_2$. Note that these results suggest that at one-loop the general energy-spin relation for these ``regular'' solutions (including the string excitations on the pp-wave)  admit an expansion in $\lambda/J \ll 1$ of the some form (\ref{expa1}),
%\bea\label{etot}
%E=m(\lambda)J\Big[1+a_1\frac{\lambda^2}{J^2}+ b_1 \frac{\lambda^2}{J_2^2} + a_2\Big(\frac{\lambda^2}{J^2}\Big)^2 + 
%b_2\Big(\frac{\lambda^2}{J_2^2}\Big)^2+\ldots\Big]\ . 
%\eea 
where $a_1,b_1,a_2,b_2,\ldots$ are again given by the classical energy-spin relation, but with $m_0$ substituted by a $\lambda$-dependent $m(\lambda)$. It seems then possible to reabsorb 1-loop quantum effects in a redefinition of the mass $m(\lambda)$ of the single hadronic constituent, leaving the hope that at lower $\lambda$ quantitative information contained in the classical result does not loose  significance  because of quantum corrections. Let us observe  that, differently from what happens in $AdS_5\times S^5$, in our case the ``regular'' solutions  admit an expansion  in even powers of $\lambda$  (up to a nontrivial dependence of $m(\lambda)$). It would be interesting to understand better this structure from the dual gauge theory point of view and if it is preserved by higher order  world-sheet  corrections. Lastly, in \cite{tomilu1} it was speculated that, while the sigma-model correction could be connected with the renormalization of the mass $m_0$ of the single constituents (but also accounting for the mean field of the other constituents), the classical string dependence in the energy-spins relation could be related to the correction in the binding energy in the chain of constituents of the hadron, due to some mixing between hadrons with different internal structure.  This would explain why in the hadron built up by only one type of scalar, and so with only one possible structure, those corrections are not present.

\vskip 2cm

\begin{center}
{\large  {\em Acknowledgments}}
\end{center}
First of all, I would like to thank the coauthors of the results presented in this paper, i.e.  F. Bigazzi, A. L. Cotrone and L. A. Pando Zayas. Thanks also to T. Mateos  and P. Smyth for useful discussions. This work is supported in part by the Federal Office for
Scientific, Technical and Cultural Affairs through the "Interuniversity
Attraction Poles Programme -- Belgian Science Policy" P5/27 and by the European Community's Human
Potential Programme under contract MRTN-CT-2004-005104 `Constituents,
fundamental forces and symmetries of the universe'.

%The style of the following references should be used in all documents.

\end{document}